\documentclass[aps,prl,reprint]{revtex4-1}

\usepackage{graphicx}
\usepackage{amsmath}
\usepackage{multirow}
\usepackage{physics}

\begin{document}

\title{Assembly and coherent control of a register of nuclear spin qubits}

\author{Katrina Barnes}
\author{Peter Battaglino}
\author{Benjamin J. Bloom}
\email[Email: ]{bbloom@atom-computing.com}
\author{Kayleigh Cassella}
\author{Robin Coxe}
\author{Nicole Crisosto}
\author{Jonathan P. King}
\author{Stanimir S. Kondov}
\author{Krish Kotru}
\email[Email: ]{krish@atom-computing.com}
\author{Stuart C. Larsen}
\author{Joseph Lauigan}
\author{Brian J. Lester}
\email[Email: ]{brian@atom-computing.com}
\author{Mickey McDonald}
\email[Email: ]{mickey@atom-computing.com}
\author{Eli Megidish}
\author{Sandeep Narayanaswami}
\author{Ciro Nishiguchi}
\author{Remy Notermans}
\author{Lucas S. Peng}
\author{Albert Ryou}
\author{Tsung-Yao Wu}
\author{Michael Yarwood}

\affiliation{Atom Computing, Inc., Berkeley, California 94710, USA}

\date{August 10, 2021}

\begin{abstract}
We introduce an optical tweezer platform for assembling and individually manipulating a two-dimensional register of nuclear spin qubits.
Each nuclear spin qubit is encoded in the ground $^{1}S_{0}$ manifold of $^{87}$Sr and is individually manipulated by site-selective addressing beams.
We observe that spin relaxation is negligible after 5 seconds, indicating that $T_1\gg5$ s.
Furthermore, utilizing simultaneous manipulation of subsets of qubits, we demonstrate significant phase coherence over the entire register, estimating $T_2^\star = \left(21\pm7\right)$ s and measuring $T_2^\text{echo}=\left(42\pm6\right)$ s.
\end{abstract}

\maketitle

The generation of a register of highly coherent, but independent, qubits is a prerequisite to performing universal quantum computation~\cite{divincenzo_2000_physical-implementation-quantum-computer}.
All operations on a quantum computer, whether quantum or classical, will suffer from errors at a rate proportional to the ratio of the length of the operation to the system's intrinsic coherence time.
Fabricated qubit systems typically embrace a technology roadmap that requires breakneck speed in both the classical and quantum control signals to realize high fidelity control~\cite{
google_2019_supremacy,
veldorst_2014_addressable-quantum-dot,
yoneda_2018_quantum-dot-spin-coherence}.
In contrast, there has been significant progress in utilizing naturally occurring quantum systems with intrinsically long-lived states to achieve the same end goal~\cite{wang_2016_single-qubit, wang_2021_hour-coherence-ion}.
However, such systems have historically struggled to achieve parallel single-site operations while maintaining the demonstrated coherence times~\cite{xia_2015_single-qubit-rb, wang_2015_coherent-addressing}.
Here we introduce a qubit encoded in two nuclear spin states of a single $^{87}$Sr atom and demonstrate coherence approaching the minute-scale within an assembled register of individually-controlled qubits.
While other systems have shown impressive coherence times through some combination of shielding, careful trapping, global operations, and dynamical decoupling~\cite{campbell_2017_optical-lattice-clock, wang_2015_coherent-addressing, wang_2021_hour-coherence-ion}, we are now able to achieve comparable -- if not longer -- coherence times while individually driving multiple qubits in parallel.
We highlight that even with simultaneous manipulation of multiple qubits within the register, we observe coherence in excess of $10^5$ times the current length of the operations, with $T_2^\text{echo} = \left(42 \pm 6\right)$ seconds.
Our results confirm that nuclear spin qubits are largely insensitive to trap induced dephasing effects without adding significant additional constraints on the trapping wavelength used.
Combined with the technical advances that have led to larger arrays of individually trapped neutral atoms~\cite{young_2020_half-minute-tweezer-clock, ebadi_2021_lukin-256-atom-simulator, scholl_2021_browaeys-quantum-simulation} and high-fidelity entangling operations~\cite{graham_2019_saffman-rydberg-gates-in-2d-array, levine_2018_lukin-rydberg-entanglement, madjarov_2020_strontium-rydberg-entanglement, levine_2019_lukin-rydberg-gates}, our results demonstrate that nuclear spin qubits offer a promising platform for the realization of intermediate-scale quantum information processors.

The proposed use of nuclear spins to encode and store quantum information has a long history owing to their isolation from undesired interactions with the environment~\cite{divincenzo_1995_quantum-computing, kane_1998_silicon-quantum-computer}.
The difficulty of reliably measuring nuclear spin states has historically limited the adoption of nuclear spin qubits outside of ensemble quantum computing demonstrations~\cite{cory_1997_ensemble-quantum-computing, gershenfeld_1997_bulk-nmr-quantum-computation}.
As the control and detection of individual quantum systems have advanced, the use of the nuclear spin degree of freedom has consistently shown favorable coherence when compared to electronic spin degrees of freedom~\cite{maurer_2012_nuclear-spin-nv-center, hensen_2020_quantum-dot-nuclear-spin-qubit, park_2017_molecular-nuclear-spin-coherence, noguchi_2011_tomography-nuclear-spin-qubit}.
However, these demonstrations have either relied on a global control architecture or tailored interactions with neighboring quantum systems, both of which impede scaling to large numbers of qubits using current technology.

%%%
% More detailed intro paragraphs:
Individually trapped neutral atoms in optical tweezers are a promising platform for the study of quantum many-body systems, combining exquisite control over the full quantum state of individual atoms with the ability to efficiently scale to larger numbers of atoms with modest overhead and minimal reduction in the per-atom fidelity~\cite{ebadi_2021_lukin-256-atom-simulator, scholl_2021_browaeys-quantum-simulation}.
Until recently, optical tweezer systems primarily used alkali metal atoms, which have favorable level structures for rapid loading and cooling of the atoms, along with ground-state hyperfine structure that enables the manipulation of metastable spin states via microwave transitions~\cite{schlosser_2001_grangier-sub-poissonian-loading, isenhower_2010_saffman-rydberg-cnot-gate, wilk_2010_browaeys-rydberg-entanglement, kaufman_2012_ground-state-cooling-tweezer, endres_2016_lukin-rearrangement}.
However, optical tweezer technology is agnostic to the specific atom chosen.
Recent work has demonstrated the ability to use the same platform for trapping alkaline-earth atoms, which have attractive properties for the storage and coherent manipulation of quantum information, as well as for cooling, state preparation, and measurement of the internal state of the atoms~\cite{norcia_2018_kaufman-strontium-in-tweezers, cooper_2018_endres-strontium-in-tweezers, saskin_2019_thompson-ytterbium-tweezers, covey_2019_2000-times-imaging, norcia_2019_kaufman-tweezer-clock, madjarov_2019_endres-strontium-clock, young_2020_half-minute-tweezer-clock}.

%--------Figure 1:--------------
\begin{figure*}[htb]
	\begin{center}
		\includegraphics[width=2.0\columnwidth]{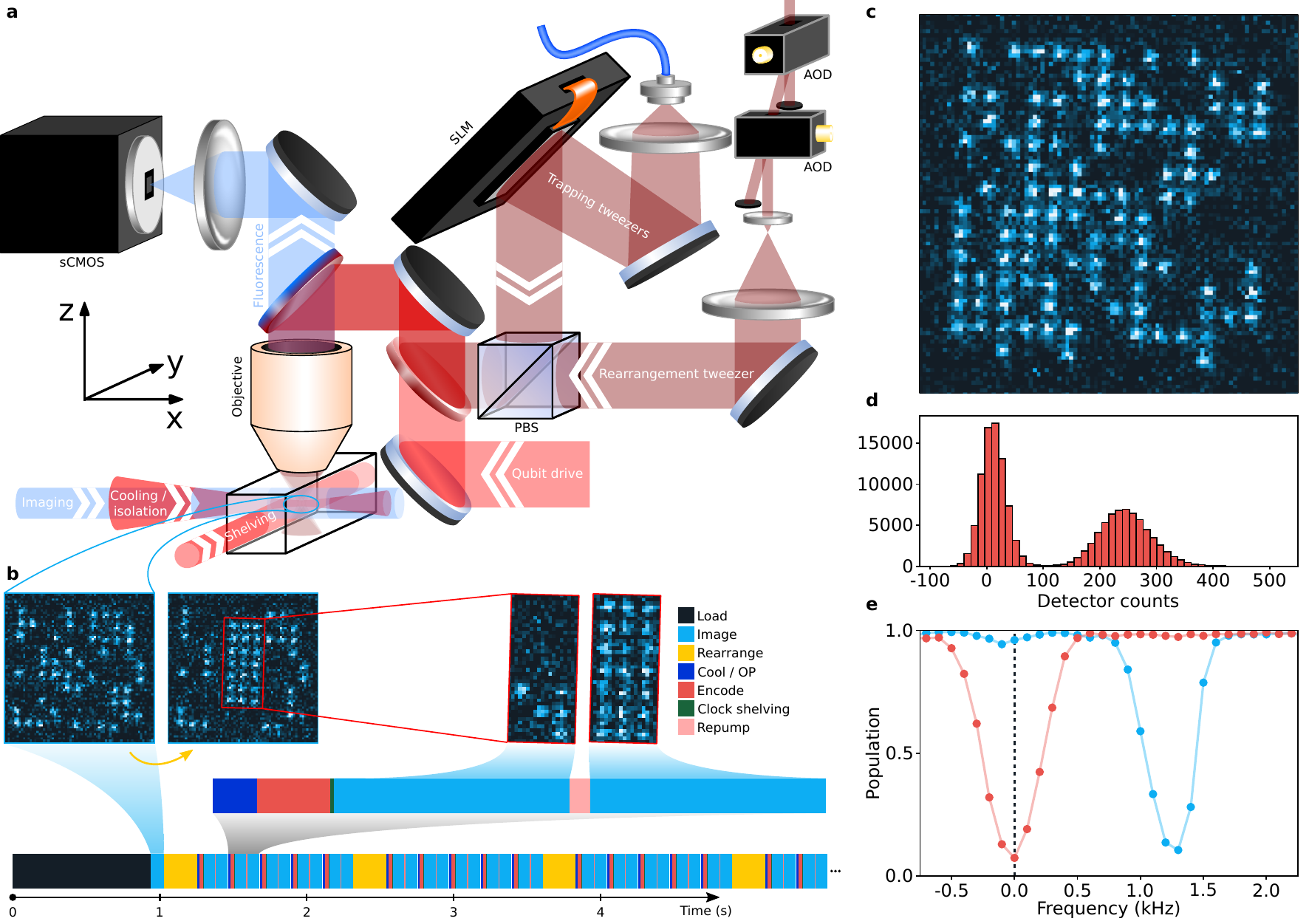}
		\caption{
			Machine to assemble a register of nuclear spin qubits.
            a) Schematic of the primary components of the optical tweezer system that trap, rearrange, manipulate, and read out the state of the nuclear spin qubits.
            The static trap array is generated using a spatial light modulator (SLM), rearrangement light is steered using a pair of crossed AODs and combined with the trap light on a polarizing beam splitter (PBS).
            Qubit drive light is combined with both of these beams using a dichroic mirror.
            All of these beams are directed into the microscope objective using another dichroic mirror that also transmits the collected atom fluorscence signal (461 nm) to the imaging system that forms an image on the commercial sCMOS camera.
            b) Experimental sequence timing. After loading the trap array, and again after several experiment repetitions (the exact number is varied based on the probability to lose an atom in each experiment), we perform rearrangement to fully fill the computational array, forming a register of qubits.
            c) Single fluorescence image demonstrating our ability to load over 100 atoms, which could be rearranged into a larger computational array than was used for this work.
            d) Histogram of photon counts collected from a single site. The two peaks indicate the presence (absence) of an atom that fluoresces on that site.
            e) Clock-state shelving spectroscopy taken when preparing the qubits in either $|\downarrow\rangle$ (red) or $|\uparrow\rangle$ (blue). By driving a transition at the frequency of the black dashed line, spin selective readout can be performed.
		}
		\label{schematic}
	\end{center}
\end{figure*}
%----------------------------------

In this manuscript we introduce Phoenix, a system for assembling a register of highly-coherent qubits encoded in the nuclear spin degree of freedom of atoms with a closed-shell $S$-orbital.
Importantly, Phoenix is able to apply tailored pulses to subsets of individual qubits in parallel---a crucial feature for gate-based quantum computation.
The system is an evolution of the alkali-based programmable quantum simulators that have recently shown great success~\cite{labuhn_2016_tunable-arrays-for-ising-simulations, bernien_2017_probing-many-body-dynamics, keesling_2019_quantum-kibble-zurek-programmable-simulator, scholl_2021_browaeys-quantum-simulation, ebadi_2021_lukin-256-atom-simulator}.
In particular, we trap individual $^{87}$Sr atoms in an array of optical tweezers, prepare a uniformly-filled register of spin-polarized atoms, then individually manipulate  and read out the spin state of the qubits.
We highlight the coherence of quantum information encoded in the ground-state nuclear spin manifold of $^{87}$Sr atoms, demonstrating the advantages of this new qubit encoding and therefore the promise of this platform for quantum information storage.

%%%
% 87Sr in optical tweezers (State prep, population imaging, shelving, and rearrangment):
Our array of optical tweezer trapping potentials is generated holographically, as shown in Figure~\ref{schematic}(a), using a liquid-crystal on silicon spatial light modulator (LCoS SLM) which imprints a spatially varying phase pattern on the beam before it is focused by a high-numerical aperture microscope objective.
We are able to programmatically define the trap geometry (array size, shape, spacing, and relative depth) using the SLM, which gives us flexibility to rapidly change the computational array geometry depending on what is required for a particular computation.
For this work, we define a rectangular trap array with 110 total trapping sites ($10\times11$ sites) with 4-$\mu$m separation between each trap center.
The trapped atoms can be used to perform many tens of state-preparation, circuit, and measurement cycles before reloading the trap array. Figure~\ref{schematic}(b) gives an example of one such cycle. The choice of array size is somewhat arbitrary; while the $10\times11$ array allows us to load $\sim$50 $^{87}$Sr atoms on average, we show in Figure~\ref{schematic}(c) that a larger $14\times14$ array can trap over 100 atoms.

Atoms are initially loaded into the optical tweezers after several laser cooling stages described in~\cite{supplement}.
We additionally perform Sisyphus cooling on the trapped atoms to bring them near the bottom of the optical tweezer potential~\cite{snigrev_2019_blatt-fast-and-dense-strontium-mot, norcia_2018_kaufman-strontium-in-tweezers, cooper_2018_endres-strontium-in-tweezers, supplement}.
The initial loading of traps is stochastic, so we define a subset of the trap array to be the computational array, which will define our register of qubits ~\cite{schlosser_2001_grangier-sub-poissonian-loading, schlosser_2002_grangier-collisional-blockade}.
The computational array comprises 21 qubits in a fully-filled $7\times3$ sub-array of traps. We uniformly fill the computational array using a dynamic optical tweezer, pictured in Figure~\ref{schematic}(a), that drags atoms from filled sites and drops them into empty sites---a process commonly known as rearrangement~\cite{barredo_2016_browaeys-rearrangement, endres_2016_lukin-rearrangement, barredo_2018_browaeys-3d-rearrangement, supplement}.
Importantly, determining if a site is occupied is accomplished by illuminating the array with light that is near-resonance with the $^{1}S_{0} \rightarrow ^{1}P_{1}$ transition, while simultaneously cooling the atoms with light near-resonance with the $^{1}S_{0} \rightarrow ^3P_{1}$ transition ~\cite{covey_2019_2000-times-imaging, supplement}.
The resulting atomic fluorescence is then imaged onto a scientific CMOS camera.

%%%
% Clock state manipulation and shelving (i.e., spin-selective detection):
Images such as the one shown in Figure~\ref{schematic}(c), combined with thresholding derived from histograms classifying counts collected per qubit, as shown in Figure~\ref{schematic}(d), reveal the spatial locations of strontium atoms in the $^{1}S_{0}$ manifold and whether or not they are fluorescing.
However, these images do not distinguish between atoms in different nuclear spin sublevels.
To detect the nuclear spin state, we use the long-lived $^{3}P_{0}$ manifold (typically used by state-of-the-art optical atomic clocks) to shelve population that we do not want to detect~\cite{norcia_2019_kaufman-tweezer-clock, madjarov_2019_endres-strontium-clock, young_2020_half-minute-tweezer-clock}.
For example, by driving a $\pi$ rotation between the $|^{1}S_{0},\,F=9/2,\,m_{F}=-9/2\rangle$ nuclear-spin ground state and $|^{3}P_{0}, \,F=9/2,\,m_{F}=-9/2\rangle$ upper clock state, we `shelve' any population in the  nuclear spin state into the clock state (see the red measurements in Figure~\ref{schematic}(e)), such that a subsequent fluorescence image will ideally only capture photons scattered from atoms that were in another nuclear spin state.
However, the lifetime of the shelved state is reduced from its natural lifetime of thousands of seconds down to of order 1 second due to the large intensity of the trap light causing Raman scattering within the triplet spin manifolds, which leads to further decay to the ground state manifold~\cite{dorscher_2018_lattice-induced-photon-scattering, norcia_2019_kaufman-tweezer-clock, madjarov_2019_endres-strontium-clock}.
This does not prevent the efficient detection of the nuclear spin state.
In~\cite{supplement} we describe how we compensate our measurements for these readout errors. In the future, a combination of more sensitive detectors, higher collection efficiency objectives, and lower scattering readout traps will ensure high-fidelity single-shot readout, for example to implement mid-circuit measurements for quantum error correction protocols. Alternatively, state specific fluorescence without shelving would sidestep this issue entirely.

%--------Figure 2:--------------
\begin{figure}[t!]
	\begin{center}
		\includegraphics[width=\columnwidth]{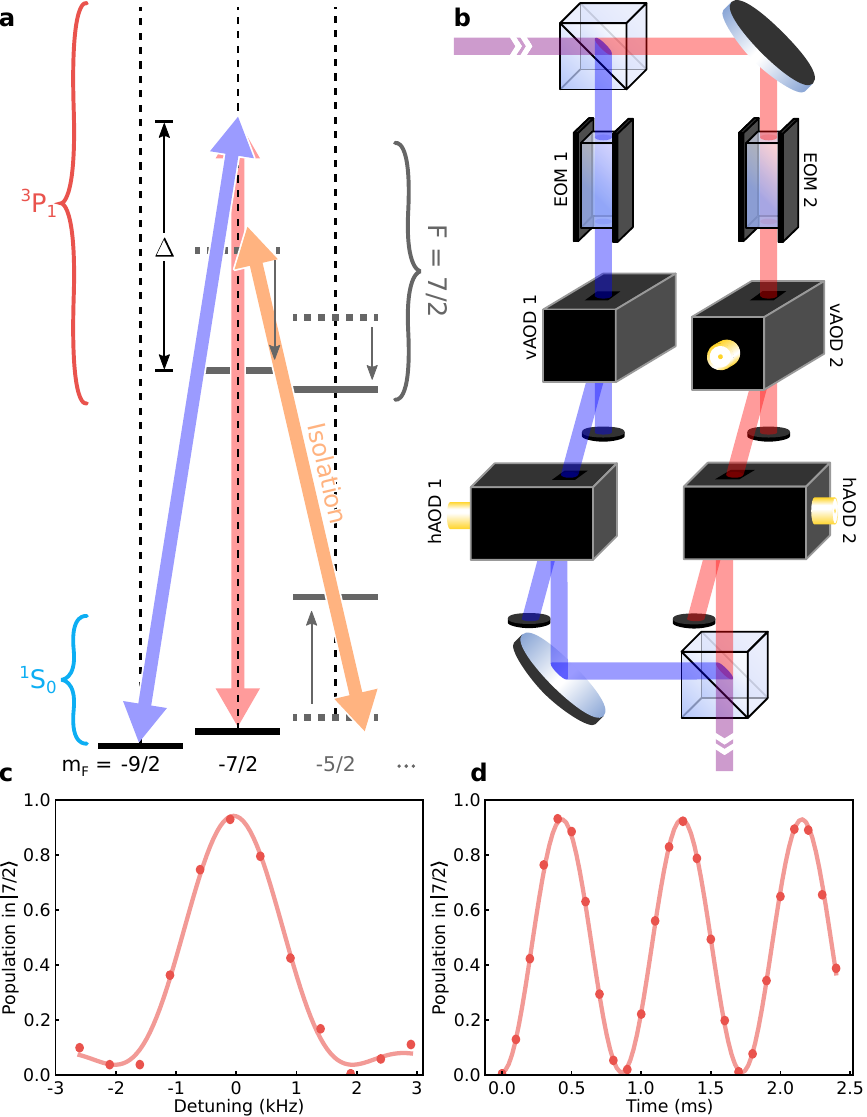}

		\caption{
			Isolating and manipulating a nuclear spin qubit.
            a) Simplified level diagram for $^{87}$Sr showing the nuclear spin qubit states $|\downarrow\rangle\equiv |^{1}S_{0},\,F=9/2,\,m_{F}=-9/2\rangle$ and $|\uparrow\rangle\equiv |^{1}S_{0},\,F=9/2,\,m_{F}=-7/2\rangle$,
            and how they are coupled via a two-photon Raman transition with two orthogonally polarized drive beams detuned by an amount $\Delta$ from the $^3P_1,\,F=7/2$ manifold.
            Also indicated is the Stark-shift beam, which is used to isolate the qubit manifold from the rest of the $I=9/2$ nuclear spin manifold by shifting the leakage transition (see main text) out of resonance with the two-photon drive.
            b) A schematic showing the preparation of the two drive beams with one electro-optic modulator (EOM) and two crossed accousto-optic deflectors (AODs) in each beam path.
            The EOMs are used for global, fast pulse shaping, while the pair of vertical and horizontal AODs (vAOD and hAOD, respectively) are used to adjust the phase and amplitude of each beam on a site-by-site basis.
            c) Nuclear spin qubit resonance measured by scanning the modulation frequency of EOM1 while driving for a fixed duration of 446 $\mu$s.
            Note that the qubit frequency is set by the applied 11-Gauss magnetic field, which defines the quantization axis.
            d) Rabi flops on a nuclear spin qubit, taken by setting the qubit frequency (which defines the exact drive frequency of EOM1) to 2.1 kHz and scanning the length of the drive pulse.
		}
		\label{nuclear_spin_qubit_figure}
	\end{center}
\end{figure}
%----------------------------------

%%%
% Qubit state definition, how to isolate:
As depicted in Figure~\ref{nuclear_spin_qubit_figure}(a), we define our qubit manifold as the two lowest-lying nuclear spin states in a positive magnetic field: $|\downarrow\rangle \equiv |^{1}S_{0},\,F=9/2,\,m_{F}=-9/2\rangle$ and
$|\uparrow\rangle \equiv |^{1}S_{0},\,F=9/2,\,m_{F}=-7/2\rangle$.
However, for any reasonably small magnetic field, the transition frequency between our qubit states will be degenerate with the leakage transition from $|\uparrow\rangle$ to $|L\rangle\equiv|^{1}S_{0},\,F=9/2,\,m_{F}=-5/2\rangle$ (and with the subsequent transitions that drive qubits further out of the qubit manifold and into higher nuclear-spin ground states).
To isolate the qubit manifold from other nuclear spin states, we apply a strong beam that is nearly resonant with the $|L\rangle\rightarrow|^{3}P_{1},\,F=7/2,\,m_{F}=-7/2\rangle$ transition, which shifts this nuclear spin state (and thus the primary leakage transition) out of resonance with any drive that performs rotations within the qubit manifold.
Because no population ends up in the $|L\rangle$ state and the polarization of the beam is such that the nearest transition from the qubit states allowed by electric dipole selection rules is far-detuned (over 1 GHz), photon scattering from the qubit states due to this beam can be suppressed to a rate of $\sim$1 Hz.

%--------Figure 3:--------------
\begin{figure}[t!]
	\begin{center}
		\includegraphics[width=\columnwidth]{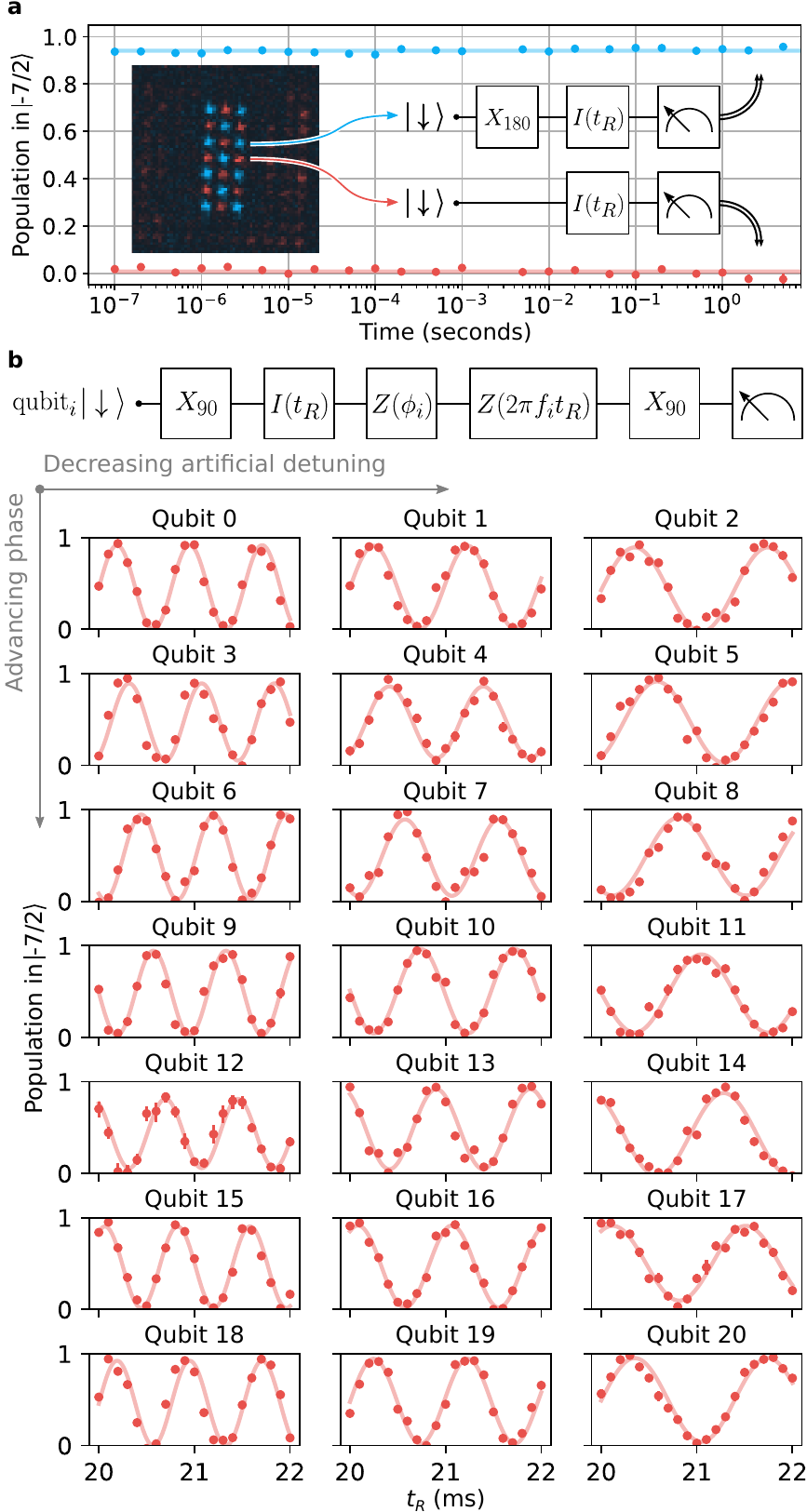}
		\caption{
            Site-resolved manipulation within the register of qubits.
            a) A measurement to bound the relaxation time ($T_1$) between the qubit states, taken in a single run by selectively rotating a subset of the qubits (indicated in the inset) to the $|\uparrow\rangle$ state before a variable hold time.
            b) A demonstration of Ramsey oscillations on individual qubits, where each qubit is given a unique combination of static phase offset $\phi_{i}$ and artificial detuning $f_{i}$ (where $f_{i}$ is the rate of phase accumulation used to set the phase of the final $\pi/2$ pulse), as specified by the circuit diagram. Solid lines are sinusoidal curves with phase offset and artificial detuning fixed to their programmed values.
            All data points have error bars representing the Wilson score confidence interval of the measurements, but in most cases these are smaller than the point markers.
		}
		\label{site_resolved_manipulation_figure}
	\end{center}
\end{figure}
%----------------------------------

%%%
% Site-resolved manipulations within the qubit manifold (Rabi):
Site-resolved qubit state manipulations are enabled by the ``Qubit drive'' beam path depicted in Figure~\ref{schematic}(a), which includes two beams that are projected through the same microscope objective. The beams are detuned from the $^{3}P_{1}$ manifold of states in order to drive two-photon Raman transitions between the nuclear spin ground states.
As shown in Figure~\ref{nuclear_spin_qubit_figure}(b), the beams share a common laser source, with each one being spatially divided and steered to the atoms using a pair of crossed acousto-optic deflectors (AODs). We achieve full amplitude and phase control over the two-photon transition at each site in the array by adjusting the radio-frequency tones driving the AODs that correspond to addressing each qubit. Importantly, the AODs are oriented such that the detuning between the beams is both finite and constant across the array of sites---this offers a separate degree of freedom for actuating the two-photon coupling, while also ensuring that atoms can be driven in parallel.
On Phoenix, we can apply operations in parallel only on atoms in a single column (or row) and serially apply drives to qubits in separate columns (or rows).
Specifically for all data presented here, the register of 21 qubits are addressed by column: This means that up to seven qubits are driven simultaneously and all 21 can be addressed with three groups of pulses.
This approach ensures that we can have full control over the operation applied to each qubit, independent of the drive applied to any other qubits.
Turning on the two-photon drive is accomplished by driving a pair of electro-optic phase modulators (EOMs), one in each of the two addressing beams (see Figure~\ref{nuclear_spin_qubit_figure}(b)). The first-order sidebands of the two EOMs drive the primary Raman process.
The EOMs enable fast pulse shaping and rapid adjustments to the intermediate state detuning.

%--------Figure 4:--------------
\begin{figure*}[t!]
	\begin{center}
		\includegraphics[width=2.0\columnwidth]{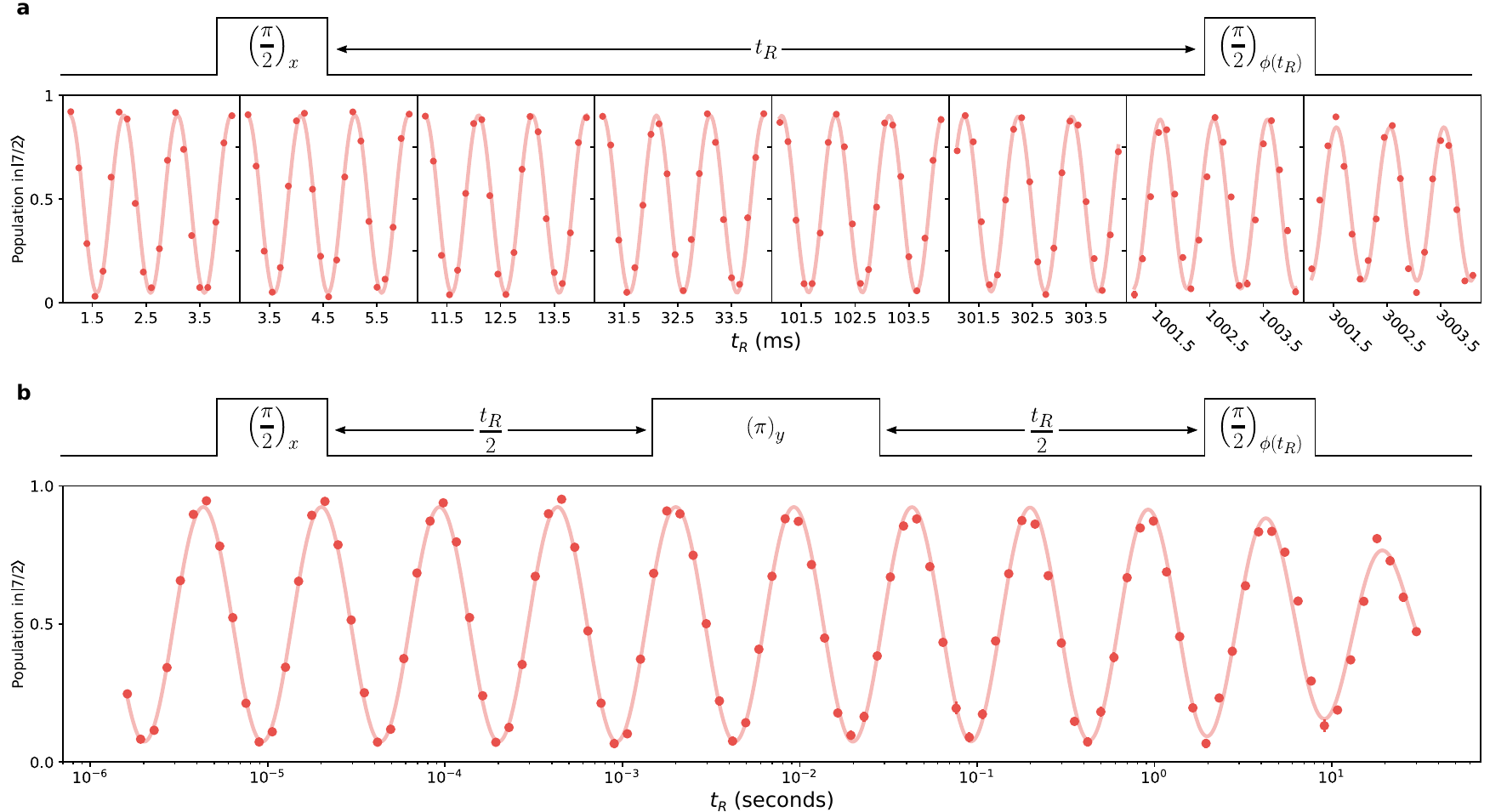}
		\caption{
			Coherence of nuclear spin qubits.
            a) Array-averaged oscillations taken in a Ramsey experiment with variable hold time, used to bound the coherence time $T_2^\star \gg 3$ s.
            The plotted exponentially decaying sinusoid is a fit to the data and gives a dephasing time $T_2^\star = \left(21 \pm 7\right)$ s.
            Note that the x-axis is split into discrete time windows of 3 ms each, spaced at exponentially increasing time offsets out to over $3$ s.
            The ability to see coherent oscillations when averaging the signal across the entire array highlights the uniformity of the qubit frequency across the array, but the slight reduction in contrast likely indicates slight miscalibrations of the pulse area used to encode and read out the phase of each qubit.
            b) Array-averaged Ramsey-echo coherence measurement taken at logarithmically spaced evolution times (as indicated in the pulse diagram, half of the evolution time happens between the first $\pi/2$ pulse and the echo $\pi$ pulse, while the remainder happens in between the echo $\pi$ pulse and the final $\pi/2$ pulse).
            By advancing the phase of the final readout pulse logarithmically in time, we can plot these measurements on a continuous semi-log plot and see oscillations that appear sinusoidal.
			The plotted curve shows an exponentially decaying sinusoid (with logarithmically advancing phase~\cite{supplement}) fitted to the data, which gives a dephasing rate $T_2^\text{echo} = \left(42 \pm 6\right)$ s.
			All data points have error bars representing the Wilson score confidence interval of the measurements, but in most cases these are smaller than the point markers.
		}
		\label{coherence_times_figure}
	\end{center}
\end{figure*}
%----------------------------------

Adjusting the drive frequency of either EOM effectively tunes the relative frequency of the two beams, which we can use to find the qubit transition, as seen in Figure~\ref{nuclear_spin_qubit_figure}(c).
By varying the length of the EOM drive, we then observe coherent Rabi oscillations between the two qubit states (see Figure~\ref{nuclear_spin_qubit_figure}(d)), demonstrating that the qubit manifold is well isolated.
Tuning the drive time to 223 $\mu$s realizes a $\pi/2$ pulse, which we define to be the $\hat{x}$-axis of the Bloch sphere.
We note that our choice of Rabi frequency of 1.16 kHz must be significantly smaller than the isolation provided by the Stark-shift beam, which is primarily limited by laser power and polarization purity.

With the ability to site-selectively drive individual qubits, we attempt to bound the spin relaxation timescale $T_1$ within the qubit manifold in a single experiment by performing standard state preparation on the full register of qubits and an additional $\pi$ rotation on 11 of the qubits (using a checkerboard pattern).
After preparing 10 qubits in $|\downarrow\rangle$ and 11 qubits in $|\uparrow\rangle$, we wait for a variable hold time to observe relaxation in the spin states over time.
As can be seen in Figure~\ref{site_resolved_manipulation_figure}(a), the depolarization timescale is significantly longer than 10 s---consistent with no depolarization on this timescale.
While this demonstrates that the qubit states are not depolarizing on timescales where we begin to be limited by the vacuum lifetime of the traps (approximately 60 s), we emphasize that increasing the vacuum lifetime beyond 400 s is routine in room-temperature atomic physics labs~\cite{covey_2019_2000-times-imaging}, and extending that further is likely possible by placing such trapping regions inside a cryogenic environment.

%%%
% Qubit coherence (Ramsey, ramsey echo):

We now turn our attention to experiments that are sensitive to the phase coherence of the qubit manifold by encoding a superposition state and reading out that superposition after some delay~\cite{ramsey_1950_original}.
The canonical experiment to demonstrate the coherence of a qubit is the Ramsey sequence, which consists of two $\pi/2$ rotations separated by a varying length of time, $t_\text{R}$, as depicted in the circuit diagram in Figure~\ref{site_resolved_manipulation_figure}(b).
In these experiments, we vary the phase of the second $\pi/2$ pulse linearly in $t_\text{R}$ to apply an ``artificial detuning'' that creates an oscillatory signal. Dephasing in the qubit manifold would generally reduce the contrast of these oscillations.

In a first Ramsey experiment, we emphasize our ability to individually drive qubits within the computational array in parallel.
In Figure~\ref{site_resolved_manipulation_figure}(b), we perform a Ramsey sequence with three unique artificial detunings $f_{i}$ on each column of seven qubits ($f_{i}\in\{0.7, 1, 1.3\}$ kHz).
Furthermore, we apply seven unique phase offsets $\phi_{i}$, one for each row of three qubits ($-\pi\leq\phi_{i}\leq\pi$).
As a result, every individual qubit should have a different Ramsey oscillation.
Note again that all seven qubits in a single column were driven simultaneously in this experiment.
The solid lines are sinusoidal fits with frequency and phase offset fixed to their programmed values, with only amplitude and vertical offset left as free parameters.
Their agreement with the data demonstrates our ability to fully control phase and frequency for every qubit.

The Ramsey oscillations are expected to decay on an exponential timescale as the qubits dephase. To measure the decay, we take similar snapshots of the Ramsey oscillations (same artificial detuning, time span, and point spacing), but with an exponentially increasing time offset.
To more clearly display these data, Figure~\ref{coherence_times_figure}(a) uses a split x-axis, where we cut out the large segments of $t_\text{R}$ that have no data present.
In contrast to Figure~\ref{site_resolved_manipulation_figure}(b), here we present the data averaged over all 21 qubits, which is possible due to the uniformity of the qubit frequency and response to the drive light across the computational array.
It is immediately clear that the contrast remains very large out to times exceeding 3 s. The solid curve is a simultaneous fit to all the measurements, highlighting the phase stability of the Ramsey oscillations.
This indicates not only that changes to the relative phase for the drive beams are small on the seconds timescale, but also that the qubit frequency is not drifting significantly on the timescale of the entire experiment (taken in pieces over the course of two days).
By fixing the frequency, phase, and initial amplitude for the fit, we can extrapolate an estimate of the dephasing timescale that is $T_2^\star = \left(21 \pm 7\right)$ s. Even without extrapolating to times longer than the plotted data, we can safely bound the dephasing as $T_2^\star \gg 3$ s.

In an attempt to more directly show the magnitude of the coherence of the nuclear spin qubits, we perform a modified Ramsey echo experiment, where we add a single ``echo'' pulse (a $\pi$ rotation about the $\hat{y}$-axis of the Bloch sphere) in between the two $\pi/2$ pulses, as depicted in Figure~\ref{coherence_times_figure}(b).
A typical Ramsey echo experiment adjusts the phase of the final $\pi/2$ pulse in the same way that the we did in the earlier Ramsey experiments. Here, we explicitly adjust the phase logarithmically in $t_\text{R}$ to give familiar sinusoidal oscillations when plotted on a semilog plot~\cite{supplement}.
The addition of the echo pulse to this sequence makes us less sensitive to any small deviation of the qubit frequency, which is important because the rate of phase accumulation at $t_\text{R}=30$ s is slow enough that even a deviation of 25 mHz ($\sim10$ ppm of the qubit frequency) would significantly alter the period of the oscillations compared to shorter $t_\text{R}$ values.
Fitting the curve with a fixed effective frequency and phase (set by fitting earlier oscillations without a decay term), and the independent variable being $\log_{10}(t_\text{R})$, we can extract an estimate of the decay time constant of $\left(42 \pm 6\right)$ s~\cite{supplement}.
While the array averaged data has been fit to estimate the overall coherence time of each qubit, the fits are consistent with the fitted values from individual site data.

In conclusion, we have demonstrated the encoding of a qubit in the nuclear spin degree of freedom of individually trapped neutral atoms.
Furthermore, we have introduced a platform that can assemble an individually-addressable register of nuclear spin qubits and is compatible with increased computational array sizes, as well as reduced gate operation times.
Future work in both strontium and other applicable elements will tackle increasing the qubit coherence time via a combination of lower noise local oscillators as well as magnetic shielding while increasing the driven Rabi rates by multiple orders of magnitude with the goal of reaching $10^8$ gates within the coherence time of the quantum system.
The ability to individually encode, manipulate, and read out these qubits is an important first step in demonstrating this platform as a leading contender for the realization of a universal quantum computer.

% \section{acknowledgments}
\begin{acknowledgments}
The authors would like to thank Sabrina Hong, Alexander Papageorge, Maxwell Parsons, Colm Ryan, and Prasahnt Sivarajah, for their contributions during the initial design and buildout of the optical systems, control hardware, and software infrastructure used in this work.
We would also like to thank Jun Ye and Alexey Gorshkov for multiple discussions during the planning stages of this work.
This work was partially supported under NSF SBIR Grant 1951188.
\end{acknowledgments}

%%%%%%%%%%%%%%%%%%%%%%
% The Bibliography is below:
%------------------------------
% \bibliography{atom_computing_bib}
% Output from bbl file below:

%%%%%%%%%%%%%%%%%%%%%%
% This command imports all the commands and text that are in the supplementary text file after the bibliography:
%%%%%%%%%% Start a new page for the supplementary text %%%%%%%%%%
\clearpage
\pagebreak
% \widetext
\begin{center}
\textbf{\large Supplementary Materials}
\end{center}
%%%%%%%%%% Merge with supplemental materials %%%%%%%%%%
%%%%%%%%%% Prefix a "S" to all equations, figures, tables and reset the counter %%%%%%%%%%
\setcounter{equation}{0}
\setcounter{figure}{0}
\setcounter{table}{0}
\setcounter{page}{1}
\makeatletter
\renewcommand{\theequation}{S\arabic{equation}}
\renewcommand{\thefigure}{S\arabic{figure}}
%%%%%%%%%% Prefix a "S" to all equations, figures, tables and reset the counter %%%%%%%%%%

\section{Atom loading, cooling, state preparation, and measurement.}
The process of initializing the qubit starts with producing a strontium atomic beam in an ultra-high vacuum (UHV) system. The atomic beam is slowed by optical forces from a Zeeman slower and 2D magneto-optical trap (MOT). A second 2D MOT then directs the atoms toward a UHV glass cell ($\sim10^{-11}$ Torr), where a 3D MOT cools an ensemble of atoms to millikelvin-scale temperatures~\cite{marty-boyd-thesis-2007}. These three cooling stages all operate on the 1S0 → 1P1 manifold transitions at 461 nm. The atoms are then further cooled by a second 3D MOT, overlapped with the 461-nm ``blue'' MOT. This second MOT operates on the narrow 1S0 → 3P1 ``intercombination'' line (natural linewidth of 7.4 kHz) at 689 nm, with laser beams that are frequency modulated to create a sawtooth-wave adiabatic passage (SWAP) MOT. The modulated light efficiently captures hotter atoms from the blue MOT~\cite{snigrev_2019_blatt-fast-and-dense-strontium-mot, norcia_2018_kaufman-strontium-in-tweezers}. When this frequency modulation stops, the narrow linewidth of the 689-nm transition is fully utilized to cool the atoms into co-located optical tweezer traps.

The tweezers operate at $\lambda$ = 813.4 nm, the magic wavelength for the optical clock transition from $|^{1}S_{0},\,F=9/2,\,m_{F}=-9/2\rangle$ to $|^{3}P_{0}, \,F=9/2,\,m_{F}=-9/2\rangle$ (Figure~\ref{schematic}(e) shows a frequency scan over this transition). We have measured radial trap frequencies of 95 kHz and trap depths of $\sim$6 MHz. The traps are loaded stochastically from the MOT cooling stages with, in some cases, $> 1$ atom. To reduce the per-trap atom number to 0 or 1, we apply a photoassociation pulse~\cite{schlosser_2002_grangier-collisional_blockade} at 461 nm such that pairs of atoms are ejected from the traps. For each trap loading cycle, we subsequently run many sequences of atom rearrangement, state-preparation, gates, and measurements (as described below; also see the sequence diagram in Figure~\ref{schematic}(b)) before reloading from MOTs becomes necessary due to atom loss during imaging or via background gas collisions. Our current vacuum-limited atom lifetime is $\sim60$ s and can be readily extended with improved pumping speeds and cryogenics.

After atom loading, which typically occurs in $< 1$ second, the individual atoms are rearranged into a uniformly filled grid near the center of the trap array using a dynamic optical tweezer controlled by the pair of crossed acousto-optic deflectors pictured in Figure~\ref{schematic}(a)~\cite{barredo_2016_browaeys-rearrangement}.
The atoms are then optically cooled to lower motional states of the trap using the Sisyphus cooling mechanism ~\cite{cooper_2018_endres-strontium-in-tweezers}.
This cooling frequency is red-detuned from the $|^{1}S_{0},\,F=9/2,\,m_{F}=-9/2\rangle\rightarrow|^{3}P_{0}, \,F=11/2,\,m_{F}=-11/2\rangle$ transition with $\sigma^{-}$ polarization, which places this process in the “attractor” regime of Sisphus cooling. Efficient cooling with a global beam is best achieved when operating with uniform-intensity traps, since variations in trap-induced light shifts across the array remain small compared to the linewidth of the cooling light. The atom temperature at the end of this stage is $\sim 4$ $\mu$K. After and during cooling, we also optically pump the atoms using light tuned near the $|^{1}S_{0}\rangle\rightarrow|^{3}P_{0}, \,F=9/2\rangle$transitions with $\sigma^{-}$ polarization.
This choice of polarization and laser detuning leaves the $|^{1}S_{0},\, m_F=-9/2\rangle$ state dark to the excitation light~\cite{takamoto_2006_katori-sr87-frequency-measurement}.

We then apply a sequence of gates to the qubits, as described in the main text, before performing a projective measurement. The measurement is performed by applying a global pulse of resonant light at 461 nm that induces atom fluorescence on the $|^{1}S_{0}\rangle\rightarrow|^{1}P_{1}\rangle$ transition, as described in the main text.
To read out the individual nuclear spin states, we “shelve” one of them into a metastable “clock state” in the $^{3}P_0$ manifold prior to applying a first imaging pulse.
In order to monitor and correct for atom loss, we post-select by repumping the shelved atoms to the ground states and then applying a second imaging pulse. The broad linewidth of the imaging transition allows for rapid photon scattering for detection, but also causes detrimental heating of the trapped atoms that can lead to atom loss. To avoid dislodging atoms from their respective tweezers, we counteract this heating by applying Sisyphus cooling simultaneously~\cite{covey_2019_2000-times-imaging}, in addition to carefully setting the intensity and frequency of the imaging light.

\section{Trap array generation and flattening}
The tweezer traps are produced at the focal plane of a custom high-NA (0.65) microscope objective with a 300-micron diffraction-limited field of view. A phase mask is imprinted on the trap light by a spatial light modulator (SLM) and optically relayed onto the back focal plane of this objective, generating nearly arbitrary and reconfigurable two-dimensional arrays of optical tweezers. The spatial phase imparted by the SLM is optically Fourier transformed by the microscope objective to create a grid of focused spots. We use the weighted Gerchberg-Saxton algorithm to calculate the appropriate phase mask for the SLM~\cite{kim_2019_Ahn-wgs-with-atoms}.

\section{Rearrangement}
Rearrangement is performed using a single focused beam derived from the same Ti:Sapphire laser source used to create the static computing traps.
This beam is steered using a pair of crossed acousto-optic deflectors (AODs), which are driven by RF waveforms generated by custom FPGA hardware.
In order to rearrange atoms into a desired target pattern, an image is first taken which establishes the locations of initially occupied sites.
This image is used to calculate a set of moves to fill target sites from the initially occupied sites according to the ``compression algorithm''~\cite{schymik2020enhanced}.
Before performing the calculated sequence of moves, the depth of the static computing traps is lowered to $\sim$20\% of their initial value.
Moves are then performed in three steps: 1. ramp up the intensity of the rearrangement beam, 2. translate the rearrangement beam from initial site to target site using linear frequency chirps on the AODs, and 3. ramp down the intensity of the rearrangement beam.

\section{Data analysis}
During readout, shelving of population in the metastable manifold of $^{3}P_0$ levels leads to photon scattering (due to the trapping light) that is detrimental to readout fidelity. We compensate for errors in reading out the $|\downarrow\rangle$ qubit level using trapped atoms outside the computational array, since they are also prepared in the $|\downarrow\rangle$ level but always undriven. Specifically, for each point in a parameter scan, we sum the counts for all the undriven atoms, conditioned on the atom remaining trapped throughout, and then normalize to the number of times an atom was present in a trap. This fractional value $p$ corresponds to the probability that an atom in the $|\downarrow\rangle$ state was measured incorrectly. We use this value to apply positive-operator value measure corrections as described in~\cite{Maciejewski_2020_Oszmaniec-POVM-correction}. In particular, each plotted measurement $m_{corr}$ is related to the raw measurement $m$ as follows:
$m_{corr} = (m - p)/(1 - p - q)$, where $q$ (the probability of incorrectly measuring the have $|\uparrow\rangle$ level) is conservatively set to 0 since it is not directly measured in the same manner as $p$.

For the Ramsey echo data in Figure~\ref{coherence_times_figure}(b), the fit function is $y = b + a e^{-t/\tau}\sin(\phi + 2\pi n \log(t))$, where $\tau$ is the decay constant, $\phi$ is a phase offset, $a$ is an amplitude, $b$ is a vertical offset, and $n$ is number of oscillations per decade of time points.

\end{document}